\documentclass{ceurart}

\usepackage{multirow}
\usepackage{hyperref}
\usepackage{subcaption}
\usepackage{float}
\usepackage[page]{appendix} 

\usepackage{listings}
\usepackage[section]{placeins}
\usepackage{caption} 
\usepackage{graphicx} 
\usepackage{todonotes}

\begin{document}

\copyrightyear{2024}
\copyrightclause{Copyright for this paper by its authors. Use permitted under Creative Commons License Attribution 4.0 International (CC BY 4.0).}
\conference{ReNeuIR 2024 (at SIGIR 2024) -- 3rd Workshop on Reaching Efficiency in Neural Information Retrieval, 18 July, 2024, Washington D.C, USA}


\title{Deep Domain Specialisation for single-model multi-domain learning to rank}

\author[1]{Abdelmaseeh Felfel}[%
email=afelfel@amazon.com
]

\author[2]{Paul Missault}[%
email=pmissaul@amazon.com
]

\begin{abstract}
Information Retrieval (IR) practitioners often train separate ranking models for different domains (geographic regions, languages, stores, websites,...) as it is believed that exclusively training on in-domain data yields the best performance when sufficient data is available. Despite their performance gains, training multiple models comes at a higher cost to train, maintain and update compared to having only a single model responsible for all domains. Our work explores consolidated ranking models that serve multiple domains. Specifically, we propose a novel architecture of Deep Domain Specialisation (DDS) to consolidate multiple domains into a single model. We compare our proposal against Deep Domain Adaptation (DDA) and a set of baseline for multi-domain models. In our experiments, DDS performed the best overall while requiring fewer parameters per domain as other baselines. We show the efficacy of our method both with offline experimentation and on a large-scale online experiment on Amazon customer traffic. 
\end{abstract}


\maketitle

\section{Introduction}

Production systems often use different ranking models for many different domains: ranking books products is not typically done with the same ranker as ranking electronics. Similarly, for a global store it is common for customers in the US to be served with a different ranker as customers in France. There are many of these \textit{domains}, and when they differ sufficiently, researchers report that training exclusively on in-domain data outperforms models that are trained on data that includes out-of-domain data \cite{missault2021addressing,chen2010knowledge}. Despite their performance gains, training multiple models comes at a higher cost to train, maintain and update compared to having only a single model responsible for all domains. 

One solution to this increased maintenance cost is to create multi-domain models. Deep Domain Adaptation (DDA) \cite{ganin2015unsupervised} is a widely popular technique to create multi-domain neural models by forcing intermediary layers to learn domain agnostic representations. Models trained with Deep Domain Adaptation learn representations that are robust across all domains, but the in-domain performance drops since the robust representations can not leverage domain-specific features and patterns \cite{chattopadhyay2020learning}.

With the recent transition from tree-based to neural ranking models \cite{qin2020neural,buyl2023rankformer} the idea of Deep Domain Adaptation came to mind as a method to create multi-domain rankers. However, in this work we do not aim to create robust domain agnostic representations, instead we propose to force domain \textit{specific} representations. With this technique, we achieve the benefits of combining multiple models into a single model, while maintaining the performance of an in-domain model. We call our technique \textbf{Deep Domain Specialisation (DDS)}. 

It is important to note that we are inspired by research in domain generalisation, but our goal is different. We aspire to create a multi-domain model that performs as well as or better than multiple single-domain models, but we make no claims about out-of-domain generalisation. We hypothesise that a single network can be made equivalent to multiple networks of single-domain models while being cheaper to train, store, and maintain.

In this work we will train a DDS model and compare it with multiple baselines including a set of domain-specific baseline models. We will show our methodology can replace multiple in-domain models in terms of performance compared to the baselines, while having lower model complexity than other consolidated approaches.

\section{Related work}
The approach of \textit{domain generalisation} was introduced to address multi-domain learning in cases where not all domains had enough domain-specific data available \cite{zhou2022domain}. Early methods for domain generalisation relied on kernel methods \cite{muandet2a013domain}, but for Deep Neural Networks the idea of Deep Domain Adaptation (DDA) is a popular baseline. It was first applied in computer vision where it showed significant out-of-domain performance increases \cite{ganin2015unsupervised}. This generalisation however comes at a cost to the source domain, since some of the domain-specific representations might be beneficial for the in-domain learning \cite{chattopadhyay2020learning}. Domain generalisation is used in many fields\cite{zhou2022domain}, and has been formalised in search ranking across stores \cite{missault2021addressing}. 

Next to the generalisation benefits of \textit{domain generalisation}, we specifically call out its benefits in terms of \textit{technical debt}. A compelling reason to reduce the amount of models is the CACE principle: "changing Anything Changes Everything" \cite{sculley2015hidden}. Responding quickly to changes (deprecation of a feature, legal compliance work, bringing improved performance) becomes more difficult as the amount of models in production increases. Alongside CACE, a single model is also more efficient to store (weights are reused). Finally, inference with a single model is easier to optimise since peak usage differs geographically and across domains.

\section{Proposed models}

\subsection{Single-Domain baseline}
\label{sec:single_domain}

The baseline in-domain model combines the latest techniques of Neural information retrieval. Specifically, we follow the RankerFormer architecture which incorporates a listwise transformer component for modeling query-product pairs jointly and enabling listwise ranking\cite{buyl2023rankformer} (as opposed to pointwise ranking with listwise loss). On top of this architecture, we add text encoders based on the BERT architecture\cite{devlin2018bert} that were pretrained on the document corpus. The architecture thus consists of four components, which were trained end-to-end:

\begin{itemize}
    \item \textbf{text similarity encoders:} 
    BERT-based encoders are used to independently encode the search query and product titles. The encoder outputs are used to compute query-product similarity features.

    \item \textbf{Trunk multilayer perceptron (MLP) component:} 
A feedforward neural network takes the text similarity features along with numerical and categorical query-product features as inputs. It generates pointwise relevance scores for each query-product pair.

    \item \textbf{Listwise transformer component:} 
    A transformer-based listwise scoring architecture. Uses the previous component's pointwise scores for all products per query as input, and jointly scores all products leveraging an attention head.

    \item \textbf{Final MLP component:} Final feedforward neural network that takes both pointwise and listwise scores from the previous layers. It generates the final purchase probability score for all query-product pairs. 
    
\end{itemize}

The architecture is depicted in Figure~\ref{fig:baseline}.

\subsection{Multi-headed MLP}

The multi-headed approach trains separate scoring heads for each domain. In the base architecture of Section~\ref{sec:single_domain}, we replace the head MLP with multiple MLP components: one for each domain. Each head outputs independent scores for each domain. By learning specialized domain-specific ranking heads, this approach aims to capture the nuances of each domain. No regularisation is placed on the lower layers, and as such the representation can capture domain-specific patterns. During inference, the right domain-specific scorer is selected by multiplication with a domain indicator feature ($domain id$ in the Figure). The total score is the sum of the domain-specific predictions weighted with this domain indicator. Figure~\ref{fig:2hd} shows the two headed architecture we use in the ranking experiment for 2 domains.

\subsection{Deep Domain Adaptation}
\label{sec:dda}
Deep Domain Adaptation (DDA) adds a domain classification head to the baseline model. Specifically, an additional MLP is appended to predict the domain of an input. Since the aim is to learn domain agnostic representations, we aim to \textit{maximize} the classification loss of domain classification. Hence a Gradient Reversal layer (GRL) is placed between the domain classifier and the rest of the network. The Gradient Reversal Layer acts as an identity function during forward propagation but during back propagation it multiplies the gradient by -1. Intuitively, GRL forces lower layers to move in the direction that increases the loss, leading to representations where classification between domains becomes impossible. The goal is to learn representations in lower layers that are invariant to the different domains. The model architecture is shown in Figure~\ref{fig:DDA}.

\subsection{DDS model}
Deep Domain Specification (DDS) uses the same architecture from Section~\ref{sec:dda} but without the Gradient Reversal Layer. This causes the network to minimize it's classification loss instead of maximising it. In turn, this forces the model to learn domain specific representations that are optimised simultaneous to do a specific task and to be maximally discriminative between input domains. The Model architecture is shown in Figure~\ref{fig:DDS}. \\

\section{Experiments}

\subsection{Experiment definition}
We will experiment in a Search relevance ranking task for 2 geographically different domains: a store in the United Arab Emirates (AE) and a store in Saudi Arabia (SA). Our goal is to create a single ranking model that can serve both the AE and SA stores, while maintaining the same or better performance as the independently trained models. This experiment is our first attempt at consolidating neural ranking model.

\subsection{Datasets}

\begin{table}[h!]
\centering
\begin{tabular}{|l|l|l|l|}
\hline
 & Train & Validation & Test \\
\hline
Dates & 1 May - 31 May & 1 June - 7 June & 8 June - 15 June \\
\hline
SA \#Queries & 576,531 & 134,180 & 139,786 \\
\hline
AE \#Queries & 1,856,515 & 378,498 & 418,932 \\
\hline
\end{tabular}
\caption{Breakdown of the experiment dataset by time period and store. This data is a uniform sample of production traffic to the stores in these geographic domains}
\label{tab:dataset}
\end{table}

The experiment used a dataset spanning May to June 2023. As seen in Table~\ref{tab:dataset}, the dataset was divided into train, validation, and test sets by time. We split by time instead of randomly to create a realistic dataset where query and user behavior patterns are not stationary. We use 1 month for training, 1 week for validation, and 1 final week held out for testing. Each query contains the customer feedback for the search query (clicks, purchases, video watches,...), and all the features, keywords and product descriptions of the top 130 ranked products shown to the customer. 

\subsection{Model training}
We trained 5 models on this dataset: DDA, DDS, Two-headed MLP and 2 store specific baselines. All models were trained using the same listwise loss to optimize for the quality of the entire ranked list of products. DDA and DDS additionally had the domain (in this case the geographic location of the store) classification loss. The two-headed model used two independent but identical losses, one for each MLP component.

\section{Results}
\subsection{Offline evaluation}
We evaluated all 5 models on the held-out test set described in Table~\ref{tab:dataset}. For each store we additionally compared the 3 consolidated approaches against the store specific baseline as shown in Table~\ref{tab:offline_analysis}. On AE test data, both the two-headed model and the DDS model outperformed the baseline where the two-headed achieved a gain of 0.48\% NDCG@16\footnote{In our online setting, we show 16 items per results page, therefore we report NDCG@16} and the DDS model had a 0.51\% gain. On SA test data, the two-headed and DDS models achieved similar NDCG gains of respectively 0.23\%  and 0.22\% over the baseline.

As expected, the DDA model showed deteriorated performance compared to the domain-specific baselines on the AE and SA test sets. DDA was optimized for generalization across stores rather than Specialisation to each store. As a domain-adapted model, DDA is likely over-regularized in a way that hurts domain-specific relevance. Since our goal is to optimise performance in known domains -not to generalize to unseen domains- DDA is not a good fit for further online testing with customers despite its consolidated architecture.

These offline results demonstrate the potential for consolidated models to not only save maintenance cost but also improve relevance over independent domain-specific models. By training on a larger dataset combined across stores, the consolidated models appear to leverage that data to perform better on each store.
 
\begin{table}[h]
\centering
\begin{tabular}{|c|c|c|}
\hline
store & Model & NDCG@16  (gain over store model) \\ \hline
\multirow{3}{*}{AE} & AE trained   & 0.535   \\ \cline{2-3} 
                    & DDA & 0.492 (-7.98\%)    \\ \cline{2-3} 
                    & two-headed & 0.537 (0.48\%)    \\ \cline{2-3} 
                    & \textbf{DDS} & \textbf{0.538 (0.51\%)}\\ \hline
\multirow{3}{*}{SA} & SA trained   &  0.524   \\ \cline{2-3} 
                    & DDA & 0.475 (-9.36\%)    \\ \cline{2-3} 
                    & \textbf{two-headed} & \textbf{0.526 (0.23\%)}  \\ \cline{2-3} 
                    & DDS   &  0.525 (0.22\%) \\ \hline
\end{tabular}
\caption{Offline analysis on SA and AE dataset. }
\label{tab:offline_analysis}
\end{table}

The two-headed and DDS models both increase performance compared to the baseline. Although, we can not conclude that a one is better than the other, see more details in Appendix~\ref{app:offline_evaluation}. However, DDS achieves this performance with the least model parameters: DDS uses 1 scoring head (+ 1 classification head that can be cut off after training) against the N scoring heads for an N-headed solution. 

\section{Online experimentation}
After seeing positive gains in offline evaluation, we launched an interleaving experiment\cite{Bi2022} to validate the performance of the consolidated models online. We analyzed a range of online metrics, but we report the increases in sales as interleaving credit, which has high correlation to corresponding metric from an A/B test \cite{Bi2022}. The gain is reported against the production model, which are store-specific models. As such that results of the store-specific models can be seen as "refreshes" of the production models. The experiment was run for 2 weeks with the online results shown in Table~\ref{tab:online_analysis}

\begin{table}[h]
\centering
\begin{tabular}{|c|c|c|}
\hline
store & Model & Interleaving credit  \\ \hline
\multirow{3}{*}{AE} & AE trained   & 9.15\%; p=0.0000 	   \\ \cline{2-3} 
                    & two-headed & 12.92\%; p=0.0000    \\ \cline{2-3} 
                    & \textbf{DDS} & \textbf{14.79\%; p=0.0000}  \\ \hline
\multirow{3}{*}{SA} & SA trained   &  12.91\%; p=0.0034    \\ \cline{2-3} 
                    & \textbf{two-headed} & \textbf{28.35\%; p=0.0000}  \\ \cline{2-3} 
                    & DDS &  24.67\%; p=0.0000   \\ \hline
\end{tabular}
\caption{Online Interleaving experiment results on SA and AE stores.}
\label{tab:online_analysis}
\end{table}

The online results further validate our offline findings that a single consolidated models can replace multiple in-domain models. Additionally, we see that consolidation improved performance in this experiment. Specifically, both the DDS and two-headed consolidated models outperform the production model and the store specific baselines in both AE and SA.

\section{Conclusion and Future Work}

Both the Deep Domain Specialisation (DDS) and two-headed models managed to performed as good as or better than the domain-specific models in both offline and online experiments. This empirically demonstrates our claim that consolidated models can be used to reduce technical debt, and do not necessarily incur performance losses as previously thought. 
Comparing between DDS and the two-headed approach, DDS had better performance online and can be scaled easier to expand to additional stores. The multi-headed approach needs N heads for N domains. DDS on the other hand does not require any architectural change and will scale to N domains with the same parameters. Scaling to multiple domains, more diverse domains is left for future work. We additionally want to investigate the performance gains we saw, as these were unexpected.  

\bibliography{biblio}

\clearpage
\begin{appendices}

\section{Model Figures}

\begin{table}[h]
\centering
\begin{tabular}{cc}
    \begin{minipage}{0.5\linewidth}
        \includegraphics[width=\linewidth]{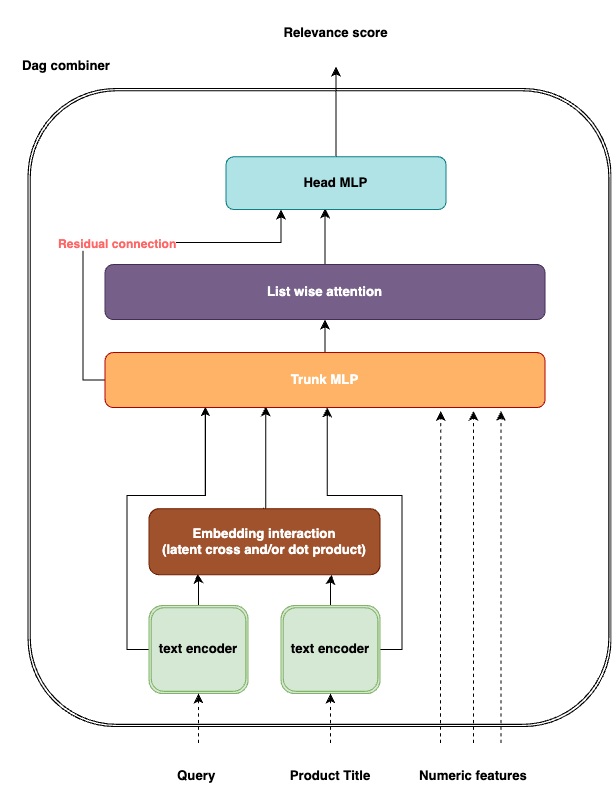}
        \captionof{figure}{Single model architecture (baseline)}
        \label{fig:baseline}
    \end{minipage}
    &
    \begin{minipage}{0.5\linewidth}
        \includegraphics[width=\linewidth]{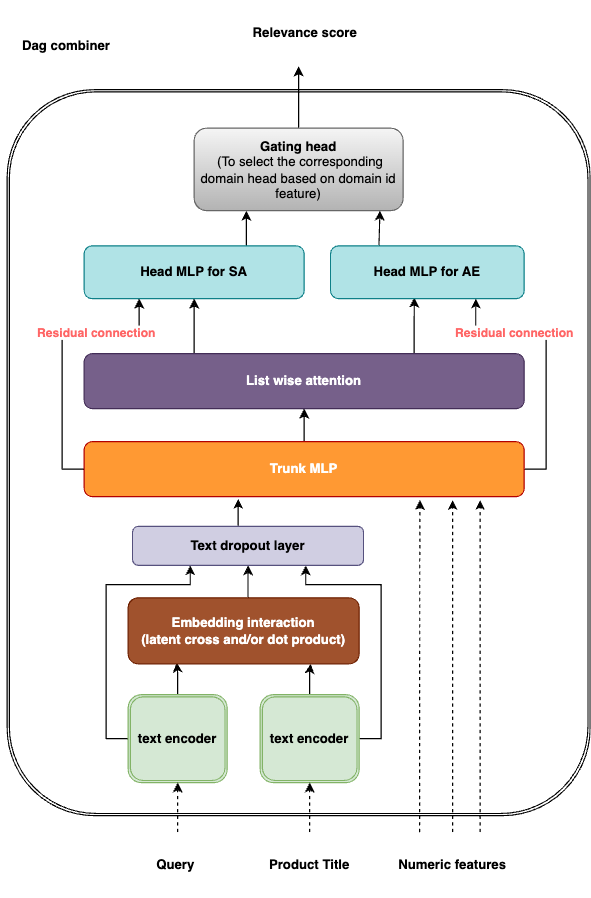}
        \captionof{figure}{Two headed architecture}
        \label{fig:2hd}
    \end{minipage}
    \\
    \begin{minipage}{0.5\linewidth}
        \includegraphics[width=\linewidth]{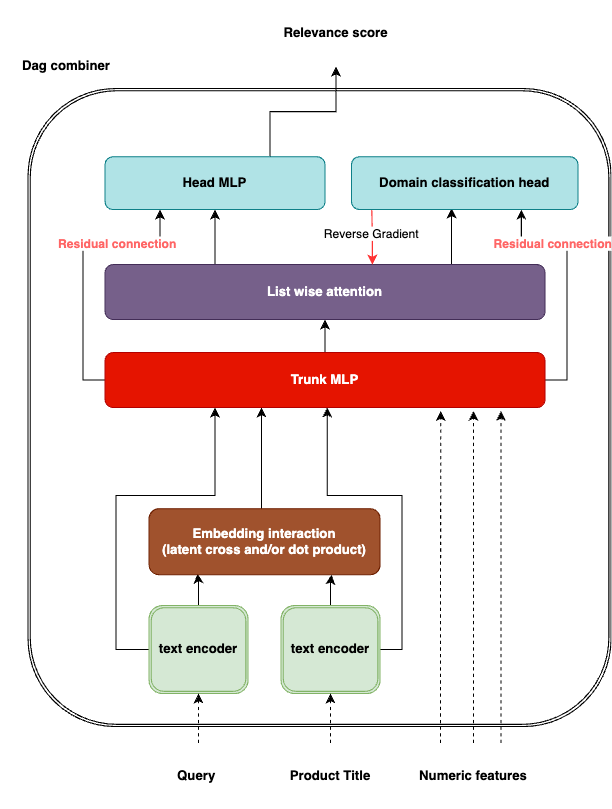}
        \captionof{figure}{DDA architecture}
        \label{fig:DDA}
    \end{minipage}
    &
    \begin{minipage}{0.5\linewidth}
        \includegraphics[width=\linewidth]{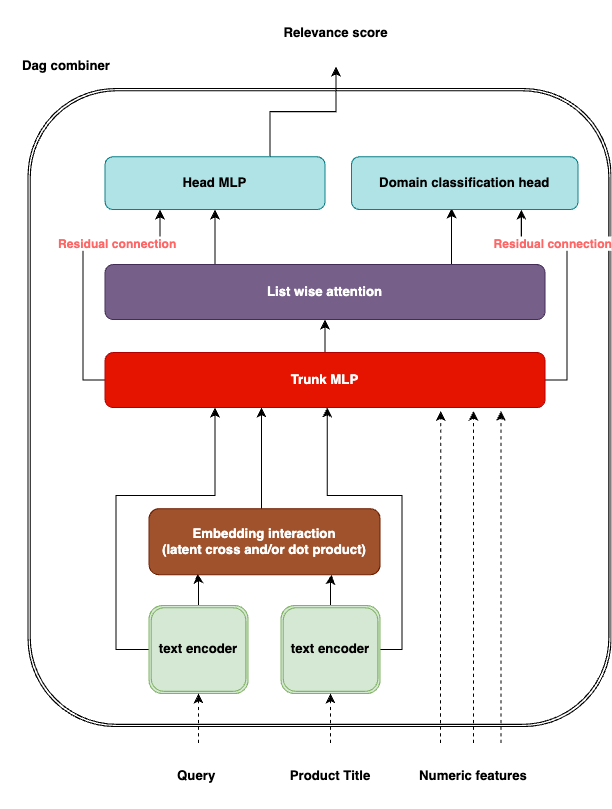}
        \captionof{figure}{DDS architecture}
        \label{fig:DDS}
    \end{minipage}
\end{tabular}
\end{table}

        




\section{Offline evaluation} \label{app:offline_evaluation}
To evaluate the models, we conducted 5 training runs for each model (the baselines, two-headed model, and DDS) with different random seeds. For each training run, we selected the model with the highest NDCG scores on a validation dataset. We then evaluated all chosen models from the 5 runs on the SA and AE test sets. Figure~\ref{fig:boxplots} below shows boxplots of the 1st and 3rd quartiles alongside the median NDCG scores across the 5 runs for each model on the two test sets.

\begin{figure}
    \centering
    \begin{subfigure}[b]{.45\linewidth}
        \centering
        \includegraphics[width=\linewidth]{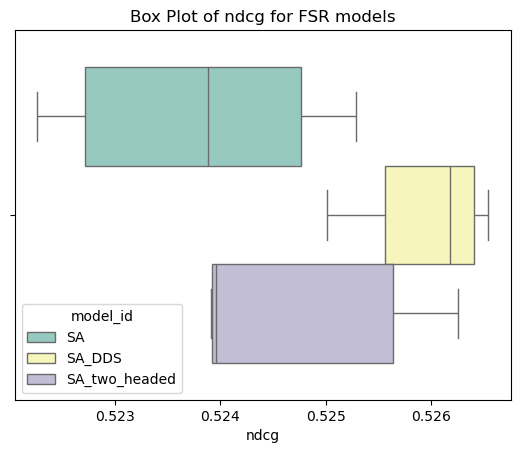}
        \caption{Models evaluation against SA test set}
      \end{subfigure}%
     \hfill
      \begin{subfigure}[b]{.45\linewidth}
        \centering
        \includegraphics[width=\textwidth]{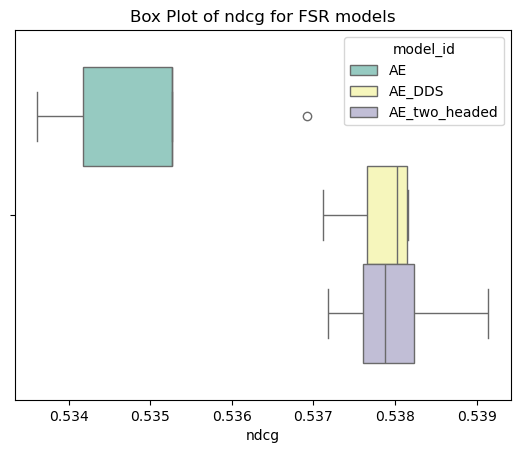}
        \caption{Models evaluation against AE test set}
        \label{fig:trafo_patten_final}
      \end{subfigure}
      \caption{NDCG scores of 5 runs of the different methods and baselines used in this paper.}
      \label{fig:boxplots}
\end{figure}

The results show that both consolidated models (DDS and multi-headed) have better or equal performance than the single domain models on both the SA and AE test sets. When comparing DDS and the two-headed models, the quartiles overlap and so we cannot conclusively determine if one model performs better than the other. As highlighted in our work, we propose DDS based on its better scaling in terms of model parameters.

\end{appendices}

\end{document}